\documentclass{eas}
\usepackage{graphicx}
\begin{document}
\title{Betelgeuse and the red supergiants}
\author{Jacco Th.\ van Loon}
\address{Lennard-Jones Laboratories, Keele University, ST5 5BG, United Kingdom}
\begin{abstract}
Betelgeuse is one of the most magnificent stars in the sky, and one of the
nearest red supergiants. Astronomers gathered in Paris in the Autumn of 2012
to decide what we know about its structure, behaviour, and past and future
evolution, and how to place this in the general context of the class of red
supergiants. Here I reflect on the discussions and propose a synthesis of the
presented evidence. I believe that, in those four days, we have achieved to
solve a few riddles.
\end{abstract}
\maketitle
\section{Is Betelgeuse typical?}

Yes. In the absence of evidence for it to be atypical, we must consider
Betelgeuse to be typical for a red supergiant. But what is `typical'? Georges
Meynet, early in the meeting, stated that ``any well observed star is a
peculiar star.'' But once we understand it, the peculiarity vanishes. Stars
vary in their evolution not just because of differences in birth mass, but
also because of differences in composition, rotation, binarity, et cetera. We
must judge how typical or peculiar Betelgeuse is, by comparison with other
stars with the same birth mass, composition, rotation, binarity, et cetera.
Because red supergiants are rare, and generally fairly distant, we lack
samples with all of the above properties known. We thus need to widen our
comparison to include stars that may be similar to Betelgeuse's past or
future, circumstantial evidence such as imprinted in the interstellar medium,
and use our fundamental understanding of physics. We could turn to models.

I would argue that on the basis of the arguments presented at this meeting,
Betelgeuse appears and behaves as expected for a star of this kind and
history.

\section{When will Betelgeuse perish -- and what will it be like?}

At least in the British media it has been claimed that Betelgeuse may explode
any moment now, and this has stuck in the minds of the public. (``If you say
it often enough it becomes reality.'' -- Ben Davies) But how correct is this?

\subsection{What is the mass of Betelgeuse?}

Betelgeuse's fate depends to a large degree on its birth mass. If sufficiently
massive, it will end as a core-collapse supernova leaving a neutron star or
black hole. But do we know that it is? If Betelgeuse were to be a lower-mass
super-Asymptotic Giant Branch (super-AGB) star it would end up as an
electron-capture supernova or die a less spectacular death leaving behind an
oxygen-neon-magnesium white dwarf possibly surrounded by a planetary nebula.
Often, Betelgeuse is assigned a birth mass of 15 M$_\odot$, which would place
it above the mass range for super-AGB stars. The mass is estimated from a
comparison between the measured effective temperature and bolometric
luminosity, $\log T_{\rm eff}({\rm K})\approx 3.56$ and
$\log L_{\rm bol}({\rm L}_\odot)\approx 5.1$, and stellar evolution models. It
thus depends on the accuracy of the model, as well as the measured distance.

Because Betelgeuse is relatively nearby one might not have thought that its
distance would be a matter of contention. Yet the distances, based on parallax
measurements, vary between 150--250 pc, corresponding to 0.4 dex in
luminosity. The stellar evolution models predict that stars with birth masses
$>15$ M$_\odot$ evolve on horizontal tracks through the Hertzsprung--Russell
diagram as they approach (and potentially leave behind) the realm of the red
supergiants. Stars with birth masses $<15$ M$_\odot$ evolve along a red
supergiant {\em branch} which takes the star on a steep path towards higher
luminosity (and somewhat cooler temperature). These branches are more
pronounced for lower masses, and hence the luminosities at the endpoints of
tracks for a 10 and a 15 M$_\odot$ star differ little. Thus, the distance
uncertainty makes it possible for Betelgeuse to be as massive as 25 M$_\odot$
but does not exclude a mass as low as 10 M$_\odot$.

However, we shall argue below that Betelgeuse is likely to be situated near
the far end of the published distance range, and thus was at least 20 M$_\odot$
at birth. This is corroborated by arguments laid out below, which suggest that
Betelgeuse is not high up a red supergiant branch. We can therefore exclude a
super-AGB nature and expect Betelgeuse to explode. But when, and what kind of
supernova?

\subsection{Betelgeuse's Dervish' dance}

Red supergiants are not supposed to rotate. This erroneous assumption has its
origin in the fact that they cannot have evolved from (much smaller) stars
rotating faster than break-up speed. Hence one would not expect red
supergiants to rotate much faster than several kilometres per second (at the
equator). Often this is considered unimportant for the behaviour of the star.
That is wrong -- speeds in the convection and pulsation movements and in the
slow winds from these cool stars are only $\sim10$ km s$^{-1}$. Also, the
detection of rotation of a red supergiant points at a history affected by it;
stellar evolution models predict differences in timescales as well as ultimate
fate, depending on the amount of rotation.

Andrea Dupree advocates that Betelgeuse rotates with $v_{\rm equator}\approx 15$
km s$^{-1}$ (Uitenbroek \etal\ \cite{uitenbroek}). This, again, would favour
Betelgeuse to have evolved from a relatively massive, large main-sequence
star. Sylvia Ekstr\"om presented the latest Geneva stellar evolution models,
that explore the effects of rotation in particular. I have extracted some
results from these tracks (Ekstr\"om \etal\ \cite{ekstrom}) for a star with a
birth mass of 20 M$_\odot$ (see Table), both without rotation and if rotating
on the main sequence at 40\% of break-up speed. We would have estimated the
age of Betelgeuse to be 8.1--8.6 Myr had we ignored rotation; it is 15\% older
if rotation is accounted for. This also meant Betelgeuse would have lost more
mass before becoming a red supergiant (RSG), hence it affects the current
mass. On the other hand, a rotating Betelgeuse will lose less mass as a red
supergiant. But more dramatically, it will not explode as a red supergiant but
first pass through a stage as a {\em blue} supergiant (BSG). The supernova
(SN) will therefore not be of type II-P, but possibly of type IIb and
remarkably similar to SN\,1987A -- which exploded as a BSG, but which exhibits
a dust ring betraying a preceding RSG phase. A rotating Betelgeuse will {\em
not} explode anytime soon.

\begin{table}
\caption{Results extracted from the Geneva models for a star with birth mass
of 20 M$_\odot$.}
\begin{tabular}{lcccc}
\hline\hline
model & RSG start & RSG duration & mass RSG start & RSG mass los\rlap{s} \\
      & (Myr)     & (Myr)        & (M$_\odot$)     & (M$_\odot$)          \\
\hline
no rotation         & 8.1 & 0.55 (then SN) & 17.3 & 8.7 \\
$v/v_{\rm crit}=0.4$ & 9.3 & 0.3 (then BSG) & 15.7 & 6.0 \\
\hline
\end{tabular}
\end{table}

\section{Betelgeuse's flight}

The position of Betelgeuse in the sky, and its movement across it, has puzzled
astronomers greatly. It is not found within any of Orion's OB associations or
star-forming nebul{\ae}. It has a large proper motion, 30 mas yr$^{-1}$, in a
direction $22^\circ$ north from east\footnote{This is also Betelgeuse's motion
with respect to the interstellar gas through which it plows, as stars formed
recently from that gas, like $\delta$\,Ori, $\epsilon$\,Ori, $\zeta$\,Ori and
25\,Ori have negligible proper motion: $\langle\mu_\alpha\rangle=1$ mas
yr$^{-1}$ and $\langle\mu_\delta\rangle=0$ mas yr$^{-1}$.}. Tracing it back, it
will have been seen in projection against the northern OB\,1b association near
the star 25\,Orionis, some 1.1 Myr ago; this association is $\sim10$ Myr old
and thus of similar age to that estimated for Betelgeuse (9.3--9.6 Myr for an
initial 20-M$_\odot$ fast-rotating star; see Table). However, the association
lies at some 250 pc from us, whilst the {\it Hipparcos} distance to Betelgeuse
is only 153 pc (a parallax of 6.55 mas; van Leeuwen \cite{vanleeuwen}).
Furthermore, the radial velocity of Betelgeuse is $v_{\rm rad}\approx 22$ km
s$^{-1}$ whilst that of 25\,Ori and most other OB stars in that region of
space is $v_{\rm rad}<20$ km s$^{-1}$, i.e.\ if anything, Betelgeuse appears to
be moving {\em towards} its potential birth place. One might follow the path
of Betelgeuse further back in the hope of finding a possible alternative place
of origin. But as its radial velocity largely reflects the Sun's peculiar
motion, Betelgeuse in fact moves pretty much perpendicular to the Galactic
plane, and towards it. Tracing it back over more than a few Myr will land it
in the Halo -- an unlikely place of origin.

How can we fix this? Graham Harper has suggested that Betelgeuse might be more
distant, around 200 pc (Harper \etal\ \cite{harper}). Convection cells render
a non-uniform brightness distribution across the 43-mas-diameter face of
Betelgeuse, causing jitter in parallax measurements that exceeds a
milli-arcsecond (Chiavassa \etal\ \cite{chiavassa2}). It thus seems possible,
and -- considering that it would solve some rather severe problems -- even
plausible, that Betelgeuse is about 250 pc distant (a true parallax of 4 mas),
and left the OB\,1a association 1.1 Myr ago, moving at a speed of $\approx 35$
km s$^{-1}$.

\subsection{Betelgeuse was a binary once}

Such a late time of ejection points at a kick velocity obtained as a result of
the explosion of a companion, rather than a result of dynamical effects within
a young compact cluster. This requires Betelgeuse to have been the less
massive component in a close binary, which is not uncommon especially among
massive stars. Tidal torques in the compact binary may have spun up the
rotation of Betelgeuse (cf.\ de Mink \etal\ \cite{selma}); this, and the
observed nitrogen enhancement in Betelgeuse's atmosphere pointed out by
Georges Meynet lend further support for both a fast rotation and a binary
origin.

Is Betelgeuse still a binary? If so, it might affect parallax measurements.
Two companions were suggested on the basis of speckle-interferometric
observations by Karovska \etal\ (\cite{karovska}), but they were never
confirmed. The proper motion of Betelgeuse is consistent with the apparent
motion of both objects in the opposite direction, hence these objects may be
real but unrelated to Betelgeuse. There is no evidence for the presence of a
compact stellar remnant -- it could be too remote from Betelgeuse to cause any
observational effects (e.g., deformation, accretion), or the binary could have
become unbound in the explosion.

\subsection{Finishing touches on Betelgeuse's deep-space sculpture}

Describing the imprint of a speedy Betelgeuse into the surrounding
interstellar medium (ISM), Jonathan Mackey likened it to ``tree rings showing
climate history.'' What can we learn from these structures about the evolution
of Betelgeuse?

Stencel \etal\ (\cite{stencel}) -- cf.\ Noriega-Crespo \etal\ (\cite{noriega})
-- discovered a double bow-shock feature ahead of Betelgeuse, in far-infrared
images obtained with the IRAS satellite. Decin \etal\ (\cite{decin}) present
exquisite images of these structures, obtained with the {\it Herschel} Space
Observatory. The off-set, near-circular inner bow is symmetric around the
proper motion vector (which differs somewhat from the assumed motion in Decin
\etal), with a radius of $\approx 0.5$ pc (for a distance to Betelgeuse of 250
pc) and a projected stand-off distance of $\approx 0.3$ pc. Decin \etal\
estimate an irradiated mass of $\sim0.003$ M$_\odot$; Le Bertre \etal\
(\cite{lebertre}) estimate $\sim0.05$ M$_\odot$ from H\,{\sc i} observations,
which is very similar to the original estimate by Noriega-Crespo \etal.
Although Orion is abound with molecular clouds, Betelgeuse is in fact moving
through a `finger' of diffuse ISM protruding from the Local Bubble (Welsh
\etal\ \cite{welsh}). These dimensions and masses thus infer a timescale of
$\sim10^{4-5}$ yr for this amount of interstellar matter and red-supergiant
wind to have been swept up, for an ISM density of $\sim1$ cm$^{-3}$ and a
mass-loss rate of Betelgeuse of $\sim3\times10^{-6}$ M$_\odot$ yr$^{-1}$ in a
wind expanding at $\sim15$ km s$^{-1}$.

Mackey \etal\ (\cite{mackey}) explained the `bar' ahead of the inner bow,
being the result of a previous, much faster blue-supergiant wind. Density
variations in the ISM may have led to the slight inclination and convex
curvature of the bar. This would not be surprising as we had seen that
Betelgeuse cannot have been a red supergiant for much more than $\sim10^5$ yr.
Mohamed \etal\ (\cite{mohamed}) suggest that the smoothness of the interface
implies an even younger age, $\sim30,000$ yr, or otherwise instabilities would
have developed. I also point out that on the ultraviolet image obtained with
the GALEX satellite, shown in Decin \etal\ (\cite{decin}), a plume of hot gas
trails Betelgeuse exactly in its wake (if the motion is as I have described
here). This is predicted in the above simulations as gas flows into the
under-pressured region behind the star, converges and heats.

\section{How Betelgeuse sheds its shroud}

Betelgeuse loses mass at a rate of 2--$4\times10^{-6}$ M$_\odot$ yr$^{-1}$,
within agreement between different mass-loss tracers: infrared re-emission of
absorbed stellar light by dust grains (Jura \& Kleinmann \cite{jura} -- but
see below), mm-wavelength emission from rotational transitions in carbon
monoxide (CO; De Beck \etal\ \cite{debeck})\footnote{Their value of
$2\times10^{-7}$ M$_\odot$ yr$^{-1}$ needs to be scaled upward by an order of
magnitude as they adopted a short distance of 131 pc and yet an exceedingly
high luminosity of $\log L({\rm L}_\odot)=5.7$.}, H\,{\i} emission (Bowers \&
Knapp \cite{bowers}), the pioneering atomic-line curve-of-growth analysis by
Weymann (\cite{weymann}), free--free radio emission (Harper \etal\
\cite{harper0}), and the bow-shock (see above). This is `exactly' what is
expected for a star of the effective temperature and luminosity of Betelgeuse
(van Loon \etal\ \cite{jacco}).

These comparisons revealed a very low dust:gas mass ratio in the outflow from
Betelgeuse, an order of magnitude below the efficient dust condensation seen
to occur in cooler red supergiants. Given the low dust content, large spread
of matter around this huge star, and modest mass-loss rate, the density in the
outflow is quite low. Consequently, the weak coupling between gas and dust
results in a rather high drift velocity of the grains with respect to the gas,
further reducing the infrared brightness of the envelope. Is this an anomaly?

No. Isabelle Cherchneff argued compellingly that an absence of water inhibits
formation of molecular clusters that would grow to become grains; while
Betelgeuse has got plenty of CO it lacks water (Tsuji \cite{tsuji}) -- though
not entirely. She further proposed that water forms in the post-shock region
of a pulsating atmosphere; Betelgeuse undergoes relatively weak pulsation,
which could explain the dearth of water, hence that of dust. Its mass loss is
likely driven not by pulsation and radiation pressure on grains, but by some
other mechanism which is -- directly or indirectly -- linked to chromospheric
activity (cf.\ Judge \& Stencel \cite{judge}).

\subsection{Resolving the basics}

Betelgeuse is near and large enough that its atmosphere and inner envelope are
spatially and kinematically resolved (Gilliland \& Dupree \cite{gilliland};
Lim \etal\ \cite{lim}; Haubois \etal\ \cite{haubois}; Kervella \etal\
\cite{kervella1,kervella2}; Ohnaka \etal\ \cite{ohnaka1,ohnaka2}). The spotty
`surface' temperature distribution leads to observed molecular patchiness, and
chaotic out- and inward motions reach $>10$ km s$^{-1}$. This points the
finger at convection to lie at the origin of the inhomogeneities (Chiavassa
\etal\ \cite{chiavassa1}), which also explains the brightness variations on
5-yr timescales (Stothers \cite{stothers}) with granulation explaining the
variability seen on 200-d timescales (Kiss \etal\ \cite{kiss}). Convection may
upset regular pulsation and thus cause the semi-regular behaviour on 400-d
timescales. Auri\`ere \etal\ (\cite{auriere}) propose that the convection
cells drive a dynamo that generates a one-Gauss magnetic field they have
detected on Betelgeuse. One may regard the convection cycle to be a heat
engine (see figure); no work is done by gravity, but work is done by the cycle
as thermal energy is transported and deposited at the base of the
chromosphere.

\begin{figure}
\includegraphics[width=12.5cm]{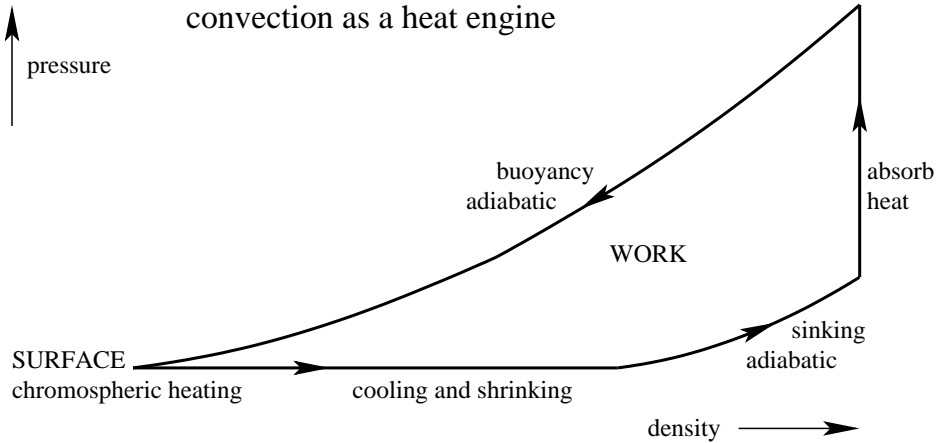}
\caption{Convection acting as a heat engine, powering the chromosphere.}
\end{figure}

We must now seek to establish a direct causal link between large upwelling
convection cells on the one hand, and discrete mass-loss events on the other;
a radio detection of non-thermal emission could signal an electro-magnetic
connection. This will require measurements to be made over a timespan of a
decade or more.

\subsection{Betelgeuse's clock is ticking -- a sign of what is to come?}

At the current rate, Betelgeuse will lose $\sim1$ M$_\odot$ over the 0.3 Myr
lifetime as a red supergiant. But as it does so, its atmosphere will cool and
expand, doubling its pulsation period (cf.\ Wood \etal\ \cite{wood}). Its
mass-loss rate must exceed $2\times10^{-5}$ M$_\odot$ yr$^{-1}$ at some point if
it were to yield the $\sim6$ M$_\odot$ predicted to be lost in the red
supergiant phase. The stronger pulsations would promote the formation of
water, molecular clusters and hence dust, and enhance the mass loss. It may
then resemble the archetypical dust-enshrouded red supergiant,
VY\,Canis\,Majoris.

VY\,CMa pulsates much more vigorously than Betelgeuse; it contains abundant
circumstellar water and molecular clusters such as TiO and TiO$_2$ (Kami\'nski
\etal\ \cite{kaminski}) which act as nucleation seeds for grain formation.
Indeed, VY\,CMa is enshrouded in a thick veil of dust, and the kinematics of
its brilliant SiO, water and hydroxyl masers trace the acceleration of the
wind through the dust-formation zone. Its mass-loss rate is vastly larger than
that of Betelgeuse; though both obey the same functional dependence on
temperature and luminosity (van Loon \etal\ \cite{jacco}), the mass-loss
mechanism of VY\,CMa is understood while that of Betelgeuse is not. Like
Betelgeuse's the envelope of VY\,CMa is far from smooth (Humphreys \etal\
\cite{humphreys}) and it will be interesting to find out whether the base of
the outflow too is as chaotic as Betelgeuse's, and whether it is for the same
reasons.

\section{Real models?}

Bernd Freytag quoted someone else saying ``all models are wrong, some models
are useful.'' As we have now seen, 3D stellar hydrodynamical models (Freytag
\etal\ \cite{freytag}) match the observed temperature distribution on
Betelgeuse rather well, even when quite a lot of the physics that we know is
important is not yet included. Maria Bergemann and Julien Lambert presented
new non-LTE model atmospheres (Bergemann \etal\ \cite{maria}), that -- in
combination with the hydrodynamical models -- should eventually permit a
meaningful comparison with flux-calibrated optical--infrared spectra (Davies
\etal\ \cite{ben3}) and the tomographic techniques described by Sophie Van Eck
(Alvarez \etal\ \cite{alvarez}; Josselin \& Plez \cite{josselin}).

\section{Populations of red supergiants: family portraits and movie snapshots}

Populations of red supergiants in different galaxies sample stellar evolution
across mass and metallicity. This provides a powerful test of the effective
temperatures and luminosities predicted by stellar evolution models (Levesque
\etal\ \cite{emily}; Drout \etal\ \cite{drout}), and their dependencies on
metallicity; it tests the predicted birth-mass range for red supergiants.
Their statistics -- assuming an initial mass function and knowledge of the
star formation history -- also yield red supergiant lifetimes (Javadi \etal\
\cite{javadi2}). Their pulsation properties offer an additional measure of
their effective gravity and hence their {\em current} mass (cf.\ Wood \etal\
\cite{wood}). Mass-loss and dust-production rates measured from infrared
surveys can be parameterised as a function of stellar parameters (van Loon
\etal\ \cite{jacco}; cf.\ Groenewegen \etal\ \cite{groen}; Bonanos \etal\
\cite{bonanos}; Mauron \& Josselin \cite{mauron}; Javadi \etal\
\cite{javadi3}), which then serve as input for the stellar evolution models.
The agreement between observations and models appears quite satisfactory --
except for the ratio of red-to-blue supergiants which is much higher than
predicted especially at lower metallicity.

Massive star clusters in the Milky Way can be used to constrain the birth-mass
range of red supergiants but also, within a single cluster, the path and speed
of evolution along the red-supergiant branch. This offers a series of
snapshots of the evolution of a star of a particular mass (within a small
margin). The limitations are that only the most massive clusters ($\sim10^4$
M$_\odot$) harbour sufficient numbers of stars to populate the realm of red
supergiants, they all have near-solar metallicity, and they are generally
heavily reddened by foreground dust in the Galactic Disc. But a handful of
populous clusters are available now, with main-sequence turn-off masses in the
range 8--20 M$_\odot$ (Davies \etal\ \cite{ben1,ben2}Clark \etal\ \cite{clark};
Negueruela \etal\ \cite{ignacio2}; Gonz\'alez-Fern\'andez \& Negueruela
\cite{gonzalez}; Negueruela \etal\ \cite{ignacio1}). The volume in the
Universe amenable to these studies will vastly expand with the advent of
groundbased 30--40m telescopes and the {\it James Webb} Space Telescope.

\section{Wishlists}

The state of modelling stellar interiors and atmospheres is truly impressive.
It is a matter of time till these models include all of the missing physics,
and they are understood well enough so we may parameterise them and
incorporate them into stellar evolution models. Hydrodynamical models will
need to incorporate rotation (differential, or solid body?) and be able to
{\em produce}, ab initio, the magnetic field, pulsation and wind. This
requires 3D atmospheres, with non-LTE feedback on the atmospheric structure
(not yet implemented) and correct treatment of patchy haze formation as well
as chromospheric activity. The resulting mixing length -- or superior
parameterisation of the effects of convection -- and mass-loss rate can be fed
into stellar evolution models. It will still be a challenge to provide the
interface between these evolutionary tracks and observables, as the meaning of
`radius' and `temperature' becomes blurred and will vary with wavelength and
time.

There is some confusion -- at least among outsiders -- as to why different
groups produce different evolutionary tracks seemingly for the same initial
properties. A perhaps more salient point that needs to be addressed is the
discrepancy between the observed ratio of blue and red supergiants, and that
predicted by the models. It would help if we could tell, from observation,
whether a blue supergiant is on its first approach towards the red-supergiant
stage or whether it is on a blue loop; likewise, if we could recognise a
second-time red supergiant; fast-evolving yellow supergiants may provide the
missing link. Direct identification of supernova progenitors, and inference
about the progenitor and circumburst medium from supernova afterglows, should
further constrain these evolutionary scenarios.

It is extremely challenging to empirically determine, with any great deal of
accuracy, the CNO and helium abundances, rotation rate, magnetic field
strength and mass-loss rate for whole populations of red supergiants. We rely
on models to extrapolate from the few nearby examples that can be studied in
great detail: Betelgeuse, its `twin' Antares (similar spectral type, similar
distance, but with a B-type companion), not forgetting the binary VV Cephei...

\section{Conclusion, for now}

We arrive at a self-consistent picture: Betelgeuse was born in the Orion
OB\,1a association at a distance of 250 pc, as a 20 M$_\odot$ close companion
to a slightly more massive star, acquiring significant rotation accompanied by
nitrogen enhancement; the primary exploded 1.1 Myr ago, furnishing Betelgeuse
with 35 km s$^{-1}$ motion through diffuse ISM. Around that same time,
Betelgeuse became a blue supergiant, and its fast wind pushed outward a dense
shell of which the infrared `bar' remains; since it became a red supergiant a
few $10^4$ yr ago, its slower wind has developed an inner, rounder shell.
Betelgeuse's pulsating convective mantle powers a magnetised chromosphere and
acoustically or Alfv\'en-wave driven wind, losing mass at a rate of
$3\times10^{-6}$ M$_\odot$ yr$^{-1}$, whilst the invigorated chemistry above
the cooler areas of receding surface layers succeeds in forming grains. As
Betelgeuse continues to expand and cool over the next couple of $10^5$ yr, the
pulsations and dust content will increase and strengthen the mass loss before
it makes a short excursion as a blue supergiant. When it blows up it will look
like SN\,1987A.

As for red supergiants, both models of stellar structure, stellar evolution
and stellar atmospheres, and observations of stellar populations in different
galaxies and rich star clusters are reaching gratifying sophistication and
detail. These models and observations nearly meet one another, but not quite
yet. Gems such as Betelgeuse remain key to marrying `seeing' and
`understanding.'

\end{document}